\documentclass[epsf]{aa}
\usepackage{natbib}
\usepackage{graphicx}
\usepackage{txfonts}
\input{aasup.sty}

\def\Porb{$P_{\rm orb}$}
\def\Psh{$P_{\rm SH}$}
\def\Pdot{$P_{\rm dot}$}
\def\Mdot{$\dot{M}$}

\begin{document}
   \title{In-the-Gap SU UMa-Type Dwarf Nova, Var73 Dra with a Supercycle
   of about 60 Days}

   \subtitle{}

   \author{Daisaku~Nogami\inst{1}
	  \and Makoto~Uemura\inst{2}
          \and Ryoko~Ishioka\inst{2}
	  \and Taichi~Kato\inst{2}
	  \and Ken'ichi~Torii\inst{3}
	  \and Donn~R.~Starkey\inst{4}
	  \and Kenji~Tanabe\inst{5}
	  \and Tonny~Vanmunster\inst{6}
	  \and Elena~P.~Pavlenko\inst{7, 8}
	  \and Vitalij~P.~Goranskij\inst{9}
	  \and Elena~A.~Barsukova\inst{10}
	  \and Oksana~Antoniuk\inst{7, 8}
	  \and Brian~Martin\inst{11}
	  \and Lewis~M.~Cook\inst{12}
	  \and Gianluca~Masi\inst{13}
	  \and Franco~Mallia\inst{14}
          }

   \offprints{D. Nogami (nogami@kwasan.kyoto-u.ac.jp)}

   \institute{Hida Observatory, Kyoto University, Kamitakara, Gifu
              506-1314, Japan
	      \and
	      Department of Astronomy, Kyoto University, Kyoto 606-8502, Japan
	      \and
	      Cosmic Radiation Laboratory, Institute of Physical and
              Chemical Research (RIKEN), 2-1, Wako, Saitama 351-0198,
              Japan
	      \and
	      DeKalb Observatory, 2507 County Road 60, Auburn, Auburn,
              Indiana 46706, USA
	      \and
	      Department of Biosphere-Geosphere Systems, Faculty of
              Informatics, Okayama University of Science, 1-1 Ridaicho,
              Okayama 700-0005, Japan
	      \and
	      Center for Backyard Astrophysics (Belgium), Walhostraat
              1A, B-3401 Landen, Belgium
	      \and
	      Crimean Astrophysical Observatory, Nauchny, 98409 Crimea,
              Ukraine
	      \and
	      Isac Newton Institute of Chile, Crimean Branch, Ukraine
	      \and
	      Sternberg Astronomical Institute, 119899, Moscow, Russia
	      \and
	      Special Astrophysical Observatory, Russian Academy of
              Sciences, Nizhnij Arkhyz, Karachaevo-Cherkesia, Russia
	      \and
	      King's University College, Department of Physics, 9125
              50th Street, Edmonton, AB T5H 2M1, Canada
	      \and
	      Center for Backyard Astrophysics (Concord), 1730 Helix
              Court, Concord, CA 94518, USA
	      \and
	      Physics Department, University of Rome "Tor Vergata" Via
              della Ricerca Scientifica, 1 00133 Rome, Italy 
	      \and
	      Campo Catino Astronomical Observatory 03025 Guarcino,
              Italy 
             }
   \date{Received ; accepted }

   \abstract{

An intensive photometric-observation campaign of the recently discovered
SU UMa-type dwarf nova, Var73 Dra was conducted from 2002 August to 2003
February.  We caught three superoutbursts in 2002 October, December and
2003 February.  The recurrence cycle of the superoutburst (supercycle)
is indicated to be $\sim$60 d, the shortest among the values known so far
in SU UMa stars and close to those of ER UMa stars.  The superhump
periods measured during the first two superoutbursts were 0.104885(93) d,
and 0.10623(16) d, respectively.  A 0.10424(3)-d periodicity was
detected in quiescence.  The change rate of the superhump period during
the second superoutburst was $1.7\times10^{-3}$, which is an order of
magnitude larger than the largest value ever known.  Outburst activity
has changed from a phase of frequent normal outbursts and infrequent
superoutbursts in 2001 to a phase of infrequent normal outbursts and
frequent superoutbursts in 2002.  Our observations are negative to an
idea that this star is an related object to ER UMa stars in terms of the
duty cycle of the superoutburst and the recurrence cycle of the normal
outburst.  However, to trace the superhump evolution throughout a
superoutburst, and from quiescence more effectively, may give a fruitful
result on this matter.

   }

   \maketitle
%
%________________________________________________________________

\section{Introduction}

Dwarf novae are a class of cataclysmic variables stars (CVs), which show
various types of variability originating in the accretion disk around
the white dwarf (for a review, \cite{war95book}).  Dwarf novae are
further classified into three basic types of SS Cyg-type dwarf novae
showing normal outbursts, Z Cam-type dwarf novae showing normal
outbursts and standstills, and SU UMa-type dwarf novae showing
superoutbursts as well as normal outbursts.  The difference of
photometric behavior in these kinds of stars including nova-like
variable stars is essentially explained by the thermal-tidal disk
instability scheme (for a review, e.g. \cite{osa96review}).  Superhumps
are oscillations with an amplitude of 0.1--0.5 mag and a period 1--5 \%
longer than the orbital period (\Porb) observed only during long, bright
(super)outbursts.  The superhump is considered to be a beat phenomenon
of the orbital motion of the secondary star and the precession of the
tidally distorted eccentric disk \citep{whi88tidal}.  The eccentricity
in such disks plays a key role to keep the accretion disk in the hot
state to make a normal outburst evolve into a superoutburst
\citep{osa89suuma}.

Non-magnetic CVs have been suggested to have a bi-modal \Porb\
distribution \citep{rob83periodgap}, while the gap between $\sim$2 h and
$\sim$3 h seems to be filled in the case of magnetic systems
\citep{web02AMHerperiodgap}.  This period gap is explained in the
standard theory of the CV evolution as follows: 1) the magnetic braking,
which is the mechanism of angular momentum loss, suddenly dies down when
the secondary star become fully convective around \Porb$\sim$3 h, 2) the
secondary shrinks into the thermal equilibrium state and the mass
transfer stops, 3) the angular-momentum loss is continued by a greatly
reduced rate by the gravitational wave radiation, and 4) the secondary
fills again its Roche-lobe around \Porb$\sim$2 h and the CV activity
restarts (for a review, \cite{kin88binaryevolution}).  Although most of
the SU UMa-type dwarf novae are distributed below the period gap, some
systems are above (TU Men: \cite{men95tumen}) and in (e.g. NY Ser:
\cite{nog98nyser}) the period gap.

The evolution scenario predicts that CVs evolve for the shorter \Porb\
region with the mass transfer rate (\Mdot) reduced, but the orbital
period begins to increase after the secondary is degenerated
\citep{pac71CVminimumperiod, kol99CVperiodminimum}.  Most SU UMa stars
are believed to be on this standard path.  However, a small group of
most active, high-\Mdot\ SU UMa stars, called ER UMa stars, has been
recently established near the period minimum \citep{kat95eruma,
nog95rzlmi, kat99erumareview}, and the evolutionary state of ER UMa
stars is a serious problem \citep{nog98CVevolution}.

Var73 Dra was discovered by \citet{ant02var73dra} on the Moscow archive
plates.  Their following CCD observations in 2001 August--October
proved that this star is an SU UMa-type dwarf nova of $R = 15.7$ at
the supermaximum and the recurrence cycle of the normal outburst is 7--8
days.  The superhump period (\Psh) was measured to be 0.0954(1) day, but
the possibility of its one-day alias, 0.1053 d, could not be rejected.

Var73 Dra is identified with USNO B1.0 1546--0228545 ($B1 = 15.90, R1 =
16.09$), the proper motion of which is not listed in the catalog.  The
SIMBAD Astronomical Database does not give any cross-identification for
this object other than the USNO entry.

We started an intensive photometric-observation campaign of Var73 Dra
since 2002 October to reveal behavior of this newly discovered
in-the-gap SU UMa-type dwarf nova.  The results including two
well-covered superoutbursts are reported in this paper.

\begin{table*}
\caption{Log of observations.}\label{tab:log}
\begin{center}
\begin{tabular}{clrcccccccc}
\hline\hline
\multicolumn{3}{c}{Date} & HJD-240000 & Exp. Time & N  & Comp.
 & Relative & filter & Instr.$^{\dagger}$ & Superhump \\
         & &             & Start--End &   (s)     &    & Star$^*$  &
 Mean Mag. &        &     \\
\hline
2002 & August    & 29 & 52516.109--52516.257 & 30  & 240 & 1 &  6.5(0.3) & no & A \\
     &           & 30 & 52517.106--52517.206 & 30  & 170 & 1 &  8.0(0.9) & no & A \\
     & September &  1 & 52519.114--52519.213 & 30  & 199 & 1 & 10.7(2.7) & no & A \\
     &           &  2 & 52520.103--52520.177 & 30  & 105 & 1 & 10.2(2.5) & no & A \\
     &           &  3 & 52520.104--52520.196 & 30  & 217 & 1 & 10.0(2.4) & no & A \\
     &           &  4 & 52522.117--52522.215 & 30  & 228 & 1 & 10.5(3.1) & no & A \\
     &           &  5 & 52523.093--52523.210 & 30  & 277 & 1 &  9.4(2.3) & no & A \\
     &           &  8 & 52526.088--52526.208 & 30  & 280 & 1 &  9.7(2.0) & no & A \\
     &           &  9 & 52527.091--52527.190 & 30  & 230 & 1 & 11.6(3.7) & no & A \\
     &           & 10 & 52528.113--52528.207 & 30  & 216 & 1 &  9.6(2.1) & no & A \\
     &           & 18 & 52536.052--52536.220 & 30  & 398 & 1 & 11.8(4.2) & no & A \\
     &           & 19 & 52537.053--52537.135 & 30  & 195 & 1 & 10.1(2.9) & no & A \\
     &           & 20 & 52538.086--52538.172 & 30  & 203 & 1 & 11.8(4.5) & no & A \\
     &           & 21 & 52539.123--52539.207 & 30  & 197 & 1 &  9.2(2.0) & no & A \\
     &           & 24 & 52542.089--52542.214 & 30  & 294 & 1 &  $>$8.0   & no & A \\
     &           & 25 & 52543.089--52543.128 & 30  & 308 & 1 & 11.2(3.7) & no & A \\
     & October   &  2 & 52550.025--52550.117 & 30  & 219 & 1 & 10.6(3.5) & no & A \\
     &           &  3 & 52551.081--52551.127 & 30  & 106 & 1 &  6.2(1.0) & no & A \\
     &           &  5 & 52553.023--52553.116 & 30  & 150 & 1 &  5.2(0.2) & no & A \\
     &           &  9 & 52557.070--52557.130 & 30  & 138 & 1 &  5.2(0.2) & no & A \\
     &           & 10 & 52557.993--52558.082 & 30  & 211 & 1 &  5.3(0.2) & no & A & $\bigcirc$ \\
     &           & 11 & 52558.993--52559.082 & 30  & 210 & 1 &  5.3(0.3) & no & A & $\bigcirc$ \\
     &           & 12 & 52559.981--52560.081 & 30  & 230 & 1 &  5.4(0.2) & no & A & $\bigcirc$ \\
     &           & 13 & 52560.674--52560.784 & 100 &  80 & 2 &  3.3(0.2) & Clear & B & $\bigcirc$ \\
     &           & 13 & 52560.917--52561.236 &  45 & 309 & 2 &  3.5(0.1) & no & C & $\bigcirc$ \\
     &           & 13 & 52560.931--52561.187 &  40 & 161 & 2 &  3.2(0.2) & no & D \\
     &           & 13 & 52560.943--52561.238 &  30 & 314 & 2 &  3.0(0.4) & no & E & $\bigcirc$ \\
     &           & 13 & 52560.944--52561.084 & 60  & 139 & 2 &  3.0(0.1) & $B$ & F & $\bigcirc$ \\
     &           & 13 & 52560.994--52561.196 &  30 & 477 & 1 &  5.9(0.3) & no & A & $\bigcirc$ \\
     &           & 14 & 52561.600--52561.762 & 120 & 107 & 8 &  1.1(0.1) & $R$ & G & $\bigcirc$ \\
     &           & 14 & 52561.885--52562.187 & 60  & 329 & 2 &  3.5(0.1) & $V$ & F & $\bigcirc$ \\
     &           & 14 & 52561.926--52562.223 &  40 & 173 & 2 &  3.3(0.4) & no  & D \\
     &           & 14 & 52561.946--52562.225 &  30 &  82 & 2 &  2.9(0.7) & no  & E \\
     &           & 14 & 52562.020--52562.228 &  55 & 151 & 2 &  3.7(0.2) & no  & C & $\bigcirc$ \\
     &           & 15 & 52562.532--52562.696 & 120 &  71 & 8 &  1.1(0.1) & $R$ & G & $\bigcirc$ \\
     &           & 15 & 52562.616--52562.722 & 60  & 145 & 2 &  3.5(0.1) & Clear & H & $\bigcirc$ \\
     &           & 15 & 52563.044--52563.230 & 30  & 436 & 1 &  6.2(0.3) & no & A \\
     &           & 15 & 52563.129--52563.266 &  55 & 118 & 2 &  3.7(0.2) & no & C & $\bigcirc$ \\
     &           & 16 & 52563.991--52564.011 & 30  &  45 & 1 &  5.8(0.2) & no & A \\
     &           & 16 & 52564.147--52564.266 & 115 &  82 & 2 &  3.7(0.3) & no & C & $\bigcirc$ \\
     &           & 17 & 52564.567--52564.666 & 150 &  52 & 8 &  1.3(0.2) & $R$ & G & $\bigcirc$ \\
     &           & 17 & 52564.979--52565.174 & 30  & 461 & 1 &  6.4(0.5) & no & A \\
     &           & 17 & 52565.077--52565.237 & 115 & 103 & 2 &  3.9(0.3) & no & C \\
     &           & 17 & 52565.401--52565.426 & 120 &  14 & 8 & 16.4(0.1)$^\ddagger$ & $R$ & K \\
     &           & 18 & 52566.202--52566.330 &  90 &  95 & 8 & 16.5(0.1)$^\ddagger$ & $R$ & K \\
     &           & 18 & 52566.256--52566.298 & 70  &  49 & 2 &  3.7(0.1) & $R$ & I & $\bigcirc$ \\
     &           & 20 & 52567.541--52567.726 & 150 &  93 & 8 &  2.4(0.2) & $R$ & G \\
     &           & 20 & 52568.236--52568.323 & 180 &  36 & 8 & 18.1(0.2)$^\ddagger$ & $R$ & K \\
     &           & 21 & 52568.979--52569.203 & 30  & 527 & 1 &  9.7(2.4) & no & A \\
     &           & 21 & 52569.191--52569.227 & 240 &   5 & 8 & 16.7(0.1)$^\ddagger$ & $R$ & K \\
     &           & 24 & 52572.221--52572.242 &  90 &  16 & 8 & 16.7(0.1)$^\ddagger$ & $R$ & K \\
     & November  & 23 & 52602.066--52602.123 & 30  & 136 & 1 & 10.2(2.7) & no & A \\
     &           & 26 & 52605.143--52605.150 & 30  &  16 & 1 &  $>$7.4   & no & A \\
     &           & 27 & 52606.088--52606.105 & 30  &  41 & 1 &  9.8(3.0) & no & A \\
     &           & 28 & 52607.036--52607.050 & 30  &  34 & 1 &  9.4(2.3) & no & A \\
     & December  &  1 & 52609.967--52609.982 & 30  &  36 & 1 &  8.3(1.4) & no & A \\
     &           &  2 & 52610.977--52610.990 & 30  &  31 & 1 &  8.9(2.1)  & no & A \\
     &           &  6 & 52614.974--52614.993 & 30  &  45 & 1 &  5.3(0.3)  & no & A \\
     &           &  8 & 52617.336--52617.447 &  80 &  97 & 3 &  4.9(0.1) & no & J & $\bigcirc$ \\
     &           &  9 & 52617.528--52617.630 & 150 &  42 & 5 &  0.4(0.1) & $R$ & G \\
     &           &  9 & 52618.173--52618.358 & 200 &  57 & 8 & 15.9(0.1)$^\ddagger$ & $R$ & K \\
     &           &  9 & 52618.241--52618.376 &  80 & 118 & 4 &  3.8(0.1) & no & J & $\bigcirc$ \\
\hline
\end{tabular}
\end{center}
\end{table*}

\setcounter{table}{0}
\begin{table*}
\caption{(continued)}\label{tab:log2}
\begin{center}
\begin{tabular}{clrcccccccc}
\hline\hline
\multicolumn{3}{c}{Date} & HJD-240000 & Exp. Time & N  & Comp.
 & Relative & filter & Instr.$^{\dagger}$ & Superhump \\
         & &             & Start--End &   (s)     &    & Star$^*$  &
 Mean Mag. &        &     \\
\hline
     &           & 10 & 52618.504--52618.557 & 150 &  30 & 6 &  0.0(0.1) & no & G & $\bigcirc$ \\
     &           & 10 & 52618.871--52619.126 &  60 & 284 & 1 &  5.6(0.3)  & no & D \\
     &           & 10 & 52618.881--52619.097 & 30  & 513 & 1 &  5.4(0.2)  & no & A \\
     &           & 10 & 52619.207--52619.394 & 80  & 175 & 7 &  4.8(0.1)  & no & J & $\bigcirc$ \\
     &           & 10 & 52619.235--52619.357 & 200 &  35 & 8 & 15.9(0.1)$^\ddagger$  & $R$ & K & $\bigcirc$ \\
     &           & 11 & 52619.856--52619.972 & 60  & 149 & 1 &  5.6(0.3)  & no & D & $\bigcirc$ \\
     &           & 11 & 52619.873--52620.099 & 30  & 507 & 1 &  5.4(0.3)  & no & A & $\bigcirc$ \\
     &           & 11 & 52620.192--52620.346 & 200 &  56 & 8 & 16.0(0.1)$^\ddagger$ & $R$ & K & $\bigcirc$ \\
     &           & 11 & 52620.243--52620.418 & 80  & 157 & 3 &  4.9(0.1)  & no & J & $\bigcirc$ \\
     &           & 12 & 52620.872--52621.085 & 60  & 264 & 1 &  5.6(0.3)  & no & D & $\bigcirc$ \\
     &           & 12 & 52620.876--52621.099 & 30  & 505 & 1 &  5.4(0.2)  & no & A & $\bigcirc$ \\
     &           & 12 & 52621.197--52621.357 & 200 &  59 & 8 & 15.9(0.1)$^\ddagger$ & $R$ & K \\
     &           & 12 & 52621.310--52621.381 & 80  &  58 & 3 &  4.9(0.1)  & no & J & $\bigcirc$ \\
     &           & 13 & 52621.882--52622.083 & 60  & 237 & 1 &  5.6(0.3)  & no & D & $\bigcirc$ \\
     &           & 14 & 52622.869--52622.970 & 30  & 232 & 1 &  5.5(0.3)  & no & A & $\bigcirc$ \\
     &           & 15 & 52623.919--52624.061 & 60  &  50 & 1 &  5.4(0.4)  & no & D \\
     &           & 15 & 52624.158--52624.268 & 120 &  58 & 8 & 16.1(0.2)$^\ddagger$ & $R$ & L \\
     &           & 16 & 52625.173--52625.297 & 180 &  45 & 8 & 16.3(0.2)$^\ddagger$ & $R$ & L \\
     &           & 17 & 52625.919--52626.051 & 60  & 162 & 1 &  6.0(0.3)  & no & D \\
     &           & 18 & 52626.142--52626.153 & 180 &   4 & 8 & 16.2(0.1)$^\ddagger$ & $R$ & L \\
     &           & 23 & 52631.892--52632.049 & 30  & 200 & 1 &  8.3(1.5)  & no & A \\
     &           & 24 & 52632.956--52633.050 & 30  &  73 & 1 &  8.7(2.2)  & no & A \\
     &           & 26 & 52634.923--52635.043 & 30  & 107 & 1 & 10.5(1.7)  & no & E \\
     &           & 27 & 52635.933--52636.038 & 30  &  56 & 1 &  $>$7.5    & no & E \\
     &           & 30 & 52638.984--52639.023 & 30  &  51 & 1 &  8.6(2.3)  & no & A \\
2003 & January   &  6 & 52646.000--52646.058 & 30  & 139 & 1 &  $>$7.4    & no & A \\
     &           &  7 & 52646.934--52647.055 & 30  & 130 & 1 &  $>$7.5    & no & A \\
     &           &  8 & 52647.933--52648.041 & 30  & 115 & 1 &  $>$7.5    & no & A \\
     &           & 11 & 52650.935--52651.038 & 30  &  73 & 1 & 5.6(0.4)   & no & A \\
     &           & 12 & 52651.982--52652.033 & 30  &  84 & 1 & 6.7(0.9)   & no & A \\
     &           & 13 & 52652.990--52653.042 & 30  &  65 & 1 & 7.3(1.7)   & no & A \\
     &           & 16 & 52656.023--52656.059 & 30  &  85 & 1 &  $>$6.4    & no & A \\
     &           & 17 & 52656.982--52657.027 & 30  &  29 & 1 &  $>$6.0    & no & A \\
     &           & 21 & 52660.894--52660.909 & 30  &  16 & 1 & 11.1(3.4)  & no & A \\
     &           & 24 & 52663.931--52663.946 & 30  &   1 & 1 &   --       & no & A \\
     &           & 28 & 52667.954--52663.959 & 30  &   2 & 1 &  $>$7.7    & no & A \\
     & February  &  4 & 52674.892--52674.898 & 30  &   9 & 1 & 6.9(0.9)   & no & A \\
     &           &  5 & 52675.941--52675.948 & 30  &   3 & 1 & 6.1(0.9)   & no & A \\
     &           &  6 & 52676.900--52676.905 & 30  &   3 & 1 &  $>$8.1    & no & A \\
     &           &  7 & 52677.903--52677.908 & 30  &   6 & 1 & 7.8(1.2)   & no & A \\
     &           & 11 & 52682.326--52682.368 & 30  &  86 & 1 & 5.3(0.1)   & no & A \\
     &           & 12 & 52683.350--52683.382 & 30  &  75 & 1 & 5.3(0.2)   & no & A \\
     &           & 13 & 52684.296--52684.384 & 30  & 209 & 1 & 5.4(0.2)   & no & A \\
     &           & 14 & 52685.290--52685.383 & 30  & 216 & 1 & 5.4(0.2)   & no & A \\
     &           & 15 & 52685.554--52685.754 & 60  & 225 & 2 & 3.3(0.1)   & no & J \\
     &           & 17 & 52687.615--52687.746 & 80  & 122 & 2 & 3.3(0.2)   & no & J \\
\hline
\multicolumn{11}{l}{$^*$1: USNO B1.0 4241--00865--1, $B1=11.48, R1=10.21$, 2: USNO B1.0 1546--0228620, $B1=13.53, R1=12.45$,} \\
\multicolumn{11}{l}{3: USNO B1.0 4241--01053--1, $B1=11.63, R1=10.91$,
 4: USNO B1.0 1545--0231804, $B1=13.37, R1=12.05$,} \\
\multicolumn{11}{l}{5: USNO B1.0 1545--0231905, $B1=16.72, R1=15.48$, 6:
 USNO B1.0 1545--0231894, $B1=17.30, R1=15.59$,} \\
\multicolumn{11}{l}{7: USNO B1.0 4241--01806--1, $B1=11.90, R1=11.45$,
 8: USNO B1.0 1545--0228537, $B1=16.51, R1=15.59$} \\
\multicolumn{11}{l}{$^{\dagger}$ A: 30-cm tel. + SBIG ST-7E (Kyoto, Japan),
 B: 12.5-inch tel. + SBIG ST-7E (Alberta, Canada), C: 30-cm tel. + SBIG
 ST-9E} \\
\multicolumn{11}{l}{(Okayama, Japan), D: 25-cm tel. + Apogee AP6E
 (Saitama, Japan), E: 20-cm tel. + Apogee AP7p (Saitama, Japan), F:
 60-cm} \\
\multicolumn{11}{l}{tel. + PixCellent S/T 00-3194 (SITe 003AB) (Hida,
 Japan), G: 36-cm tel. + SBIG ST-10XME, (Indiana, USA), H: 44-cm tel.}
 \\
\multicolumn{11}{l}{+ Genesis 16\#90 (KAF 1602e) (California, USA), I:
 80-cm tel. + SBIG  ST-9 (Campo Catino Observatory, Italy), J: 35-cm
 tel.} \\
\multicolumn{11}{l}{+ SBIG ST-7 (Landen, Belgium), K: 60-cm tel. + SBIG
 ST-7 (Crimea, Ukraine), L: 38-cm tel. + SBIG  ST-7 (Crimea, Ukraine)} \\
\multicolumn{11}{l}{$^{\ddagger}$ The magnitude is adjusted to the
 Johnson $R$ magnitude, using $R=15.58$ of the comparison star 9
 \citep{ant02var73dra}.} \\
\end{tabular}
\end{center}
\end{table*}

\section{Observations}

\begin{figure}
\centering
\includegraphics[width=8.4cm]{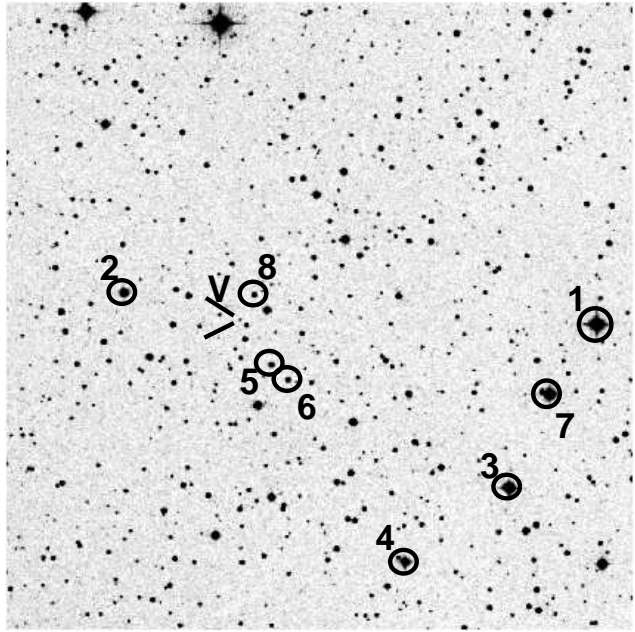}
\caption{Finding chart of Var73 Dra generated by the astronomical
 image-data server operated by National Astronomical Observatory of
 Japan, making use of Digital Sky Survey 2 (Region ID: XP106, Plate ID:
 A0LI).  North is up, and East is left.  The field of view is
 $13'\times13'$.  The numbers from 1 to 8 are given to the comparison
 stars in Table \ref{tab:log}
}
\label{fig:chart}
\end{figure}

The observations were carried out at ten sites with twelve sets of
instruments.  The log of the observations and the instruments are
summarized in Table \ref{tab:log}.  Figure \ref{fig:chart} is a finding
chart where the comparison stars are marked.

All the frames obtained at Hida and Okayama, and frames at Saitama on
2002 October 14 were reduced by the aperture photometry package in
IRAF\footnote{IRAF is distributed by the National Optical Astronomy
Observatories for Research in Astronomy, Inc. under cooperative
agreement with the National Science Foundation.}, after de-biasing (Hida
frames) or dark-subtraction (Okayama and Saitama frames), and
flat-fielding.  The Kyoto frames and the rest of the Saitama frames were
processed by the PSF photometry package developed by one of the authors
(TK).  All frames obtained at the DeKalb Observatory, CBA Belgium, and
CBA Concord were reduced by aperture photometry after dark subtraction
and flat-fielding, using the AIP4WIN software by Berry and
Burnell\footnote{http://www.willbell.com/aip/index.htm}.  The Crimean
images were dark subtracted, flat-fielded and analyzed with the
profile/aperture photometry package developed by  one of the authors
(VPG).

\section{Results}

\begin{figure}
\centering
\includegraphics[width=8.4cm]{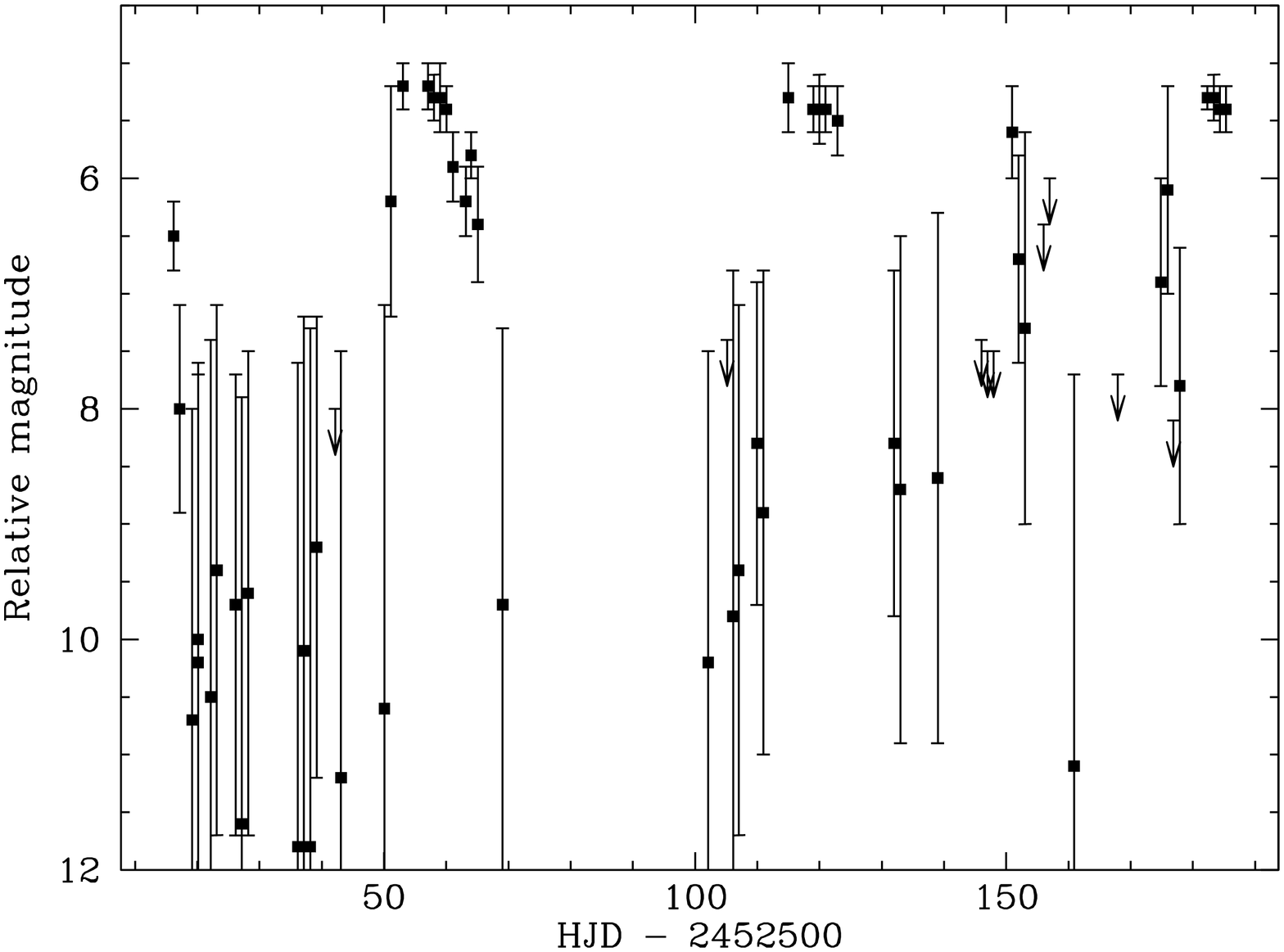}
\caption{Long-term light curve of Var73 Dra drawn with the Kyoto data
 only.  The campaign was started at the decline phase of an outburst.
 Three superoutbursts were observed around HJD 2452560, 2452620, and
 2452680.  A normal outburst was recorded around HJD 2452650.}
\label{fig:long}
\end{figure}

The long-term light curve is shown Fig. \ref{fig:long}.  During our
monitoring, Var73 Dra gave rise to three superoutbursts: the first was in
the rising phase on HJD 2452553, the second began on some day between
HJD 2452611 and HJD 2452614 (see Table \ref{tab:log}), the precursor of
the third superoutburst was caught on HJD 2552674.  This fact proves
the supercycle of Var73 Dra to be $\sim$60 days.  While two normal
outbursts were caught at the start and around HJD 2452650 in the
long-term light curve shown in Fig. \ref{fig:long}, our observations
reject the possibility of the recurrence cycle of the normal outburst
shorter than 15 days.

\begin{figure}
\centering
\includegraphics[width=8.4cm]{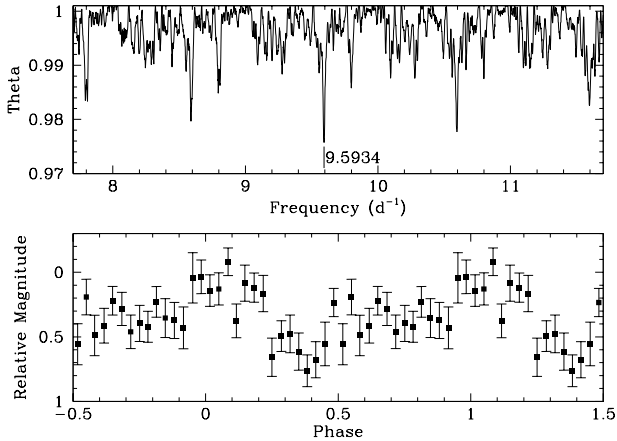}
\caption{{\bf a)} PDM Theta diagram of a period analysis of the
 quiescence data between 2002 August 30 and October 3 (see text).  A
 period of 0.10424(3) d is pointed.  {\bf b)} The quiescence light
 curve folded by the 0.10424-d period after subtracting the daily
 average magnitude from the data.}
\label{fig:q}
\end{figure}

To search periodic variability in quiescence, a period analysis by the
Phase Dispersion Minimization (PDM) method \citep{PDM} was performed for
the data obtained between 2002 August 30 and October 3, after excluding
points over 3$\sigma$ far from the daily mean magnitude and subtracting
the daily mean magnitude from the daily data sets.  Fig. \ref{fig:q}
exhibits the resultant theta diagram.  The sharp peak points to the
period of 0.10424(3) d.  The error of the period was estimated using the
Lafler-Kinman class of methods, as applied by \citet{fer89error}.  The
folded light curve have a strong peak at $\phi=0.0$ and a marginal one
around $\phi\sim0.6$.

\begin{figure}
\centering
\includegraphics[width=8.4cm]{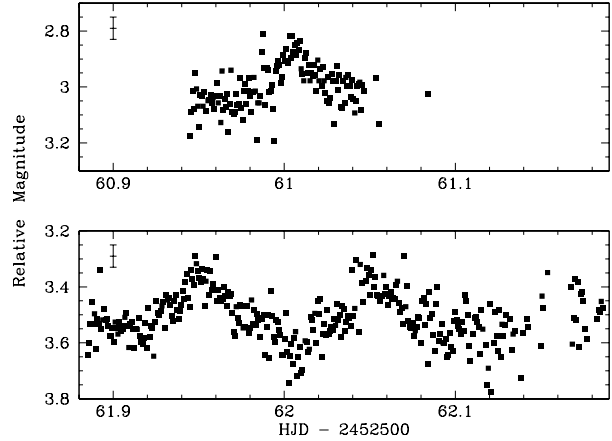}
\caption{Superhumps observed at the Hida observatory on 2002 October 13
 (upper panel) and 14 (lower panel).  The typical error bars are drawn
 near the upper-left corner.}
\label{fig:superhump}
\end{figure}

\begin{figure}
\centering
\includegraphics[width=8.4cm]{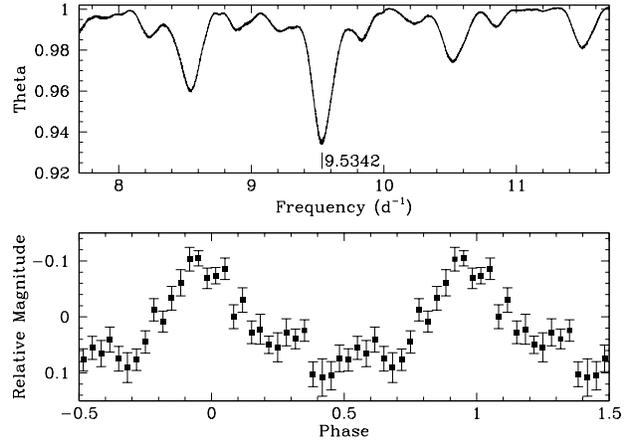}
\caption{{\bf a)} PDM theta diagram for superhumps observed during
 the first superoutburst.  The best estimated superhump period is
 0.104885(93) d.  {\bf b)} Superhump light curve folded by the
 superhump period, after subtracting the mean magnitude from each data
 set.}
\label{fig:super1}
\end{figure}

Fig. \ref{fig:superhump} shows examples of superhumps observed during
the first superoutburst.  After selecting data sets with errors small
enough to use for the period analysis (indicated by $\bigcirc$ in Table
\ref{tab:log}) and subtracting the mean magnitude from each data set, we
applied the PDM period analysis to the processed data sets.  The theta
diagram and the mean superhump light curve is given in
Fig. \ref{fig:super1}.  The superhump period of 0.104885(93) d we
obtained affirms the longer candidate proposed by \citet{ant02var73dra},
and assures that Var73 Dra is an in-the-gap SU UMa-type dwarf nova with
the second longest \Psh, next to TU Men \citep{StolzSchoembs}, almost
equal to that of NY Ser \citep{nog98nyser}.

\begin{table}
\caption{Timings of the superhump maxima during the first
 superoutburst.}\label{tab:shmax1}
\begin{center}
\begin{tabular}{crrr}
\hline\hline
HJD$-$2452500 &  E &  O$-$C1$^*$ & O$-$C2$^\dagger$ \\
\hline
61.0208(22) &  0 &  0.0012 &  0.0010 \\
61.1133(18) &  1 & $-$0.0009 & $-$0.0011 \\
61.6372(20) &  6 & $-$0.0004 & $-$0.0020 \\
62.5826(24) & 15 &  0.0028 &  0.0027 \\
62.6823(16) & 16 & $-$0.0021 &  0.0016 \\
\hline
\multicolumn{4}{l}{$^*$ Using Eq. (1).}\\
\multicolumn{4}{l}{$^\dagger$ Using Eq. (2).}\\
\end{tabular}
\end{center}
\end{table}

\begin{figure}
\centering
\includegraphics[width=8.4cm]{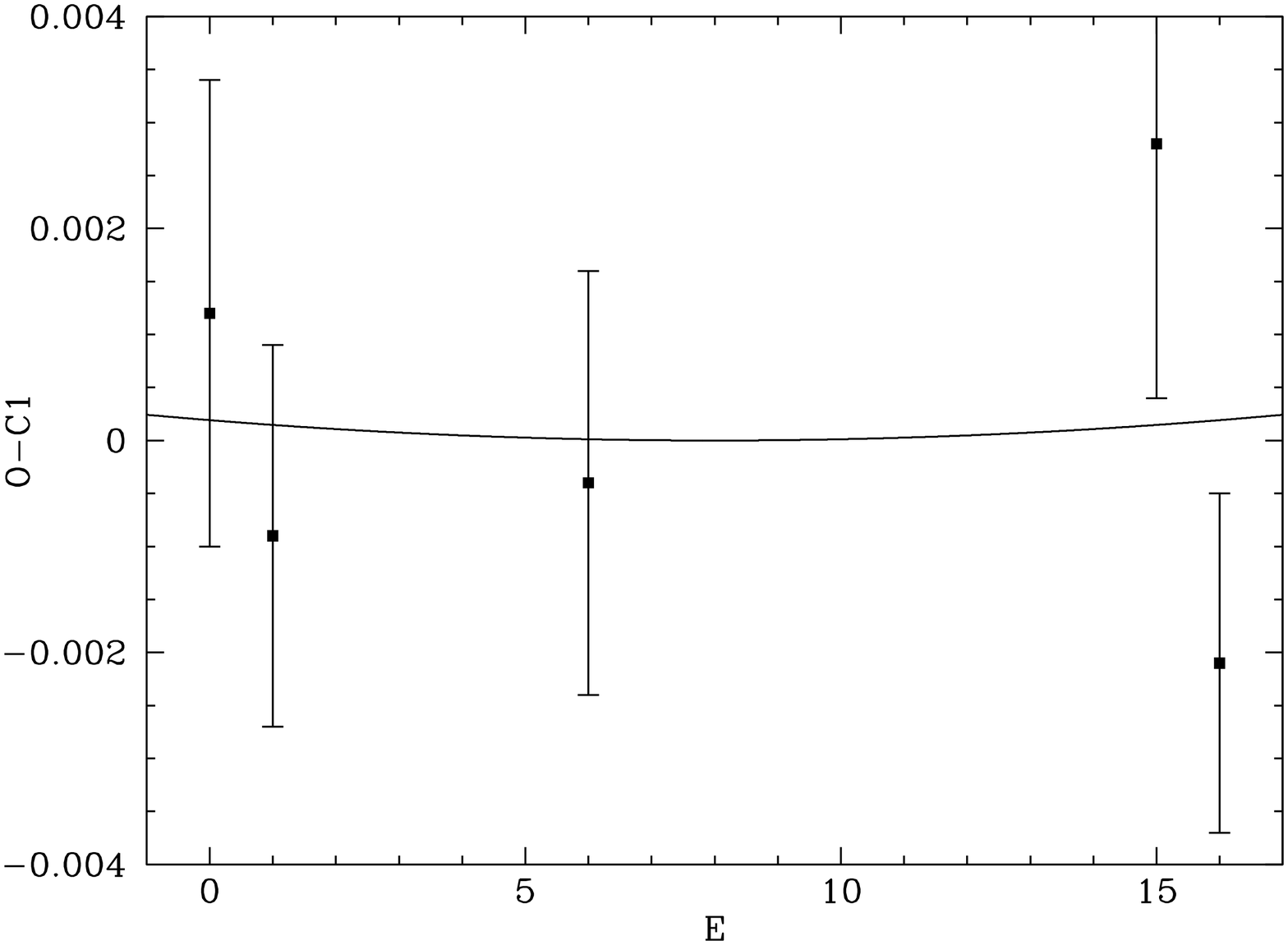}\label{fig:shmax1}
\caption{$O-C$ diagram of the timings of the superhump maxima in Table
 \ref{tab:shmax1}.  The calculated timings are given by Eq. (1).  The
 parabolic curve is based on Eq. (2).}
\end{figure}

We extracted the timings of the superhump maxima by fitting the average
superhump light curve in Fig. \ref{fig:super1}.  The results are listed
in Table \ref{tab:shmax1}.  The cycle count $E$ was set to be 0 at the
first superhump maximum measured.  A linear regression and a parabolic
fit to the times give the following equations:
\begin{equation}
 HJD_{\rm max} = 61.847(1) + 0.10468(15)\times(E-8),
\end{equation}
and
\begin{eqnarray}
 HJD_{\rm max}&=&61.847(3) + 0.10468(18)\times(E-8) \nonumber \\
              & &  + 0.000003(56)\times(E-8)^2.
\end{eqnarray}
The ordinate of Fig. \ref{fig:shmax1} represents the deviation of the
observed timing from the expected one by Eq. (1), $O-C1$, and the curve
is drawn based on Eq. (2).  We could not significantly determine the
change rate of the superhump period.

\begin{figure}
\centering
\includegraphics[width=8.4cm]{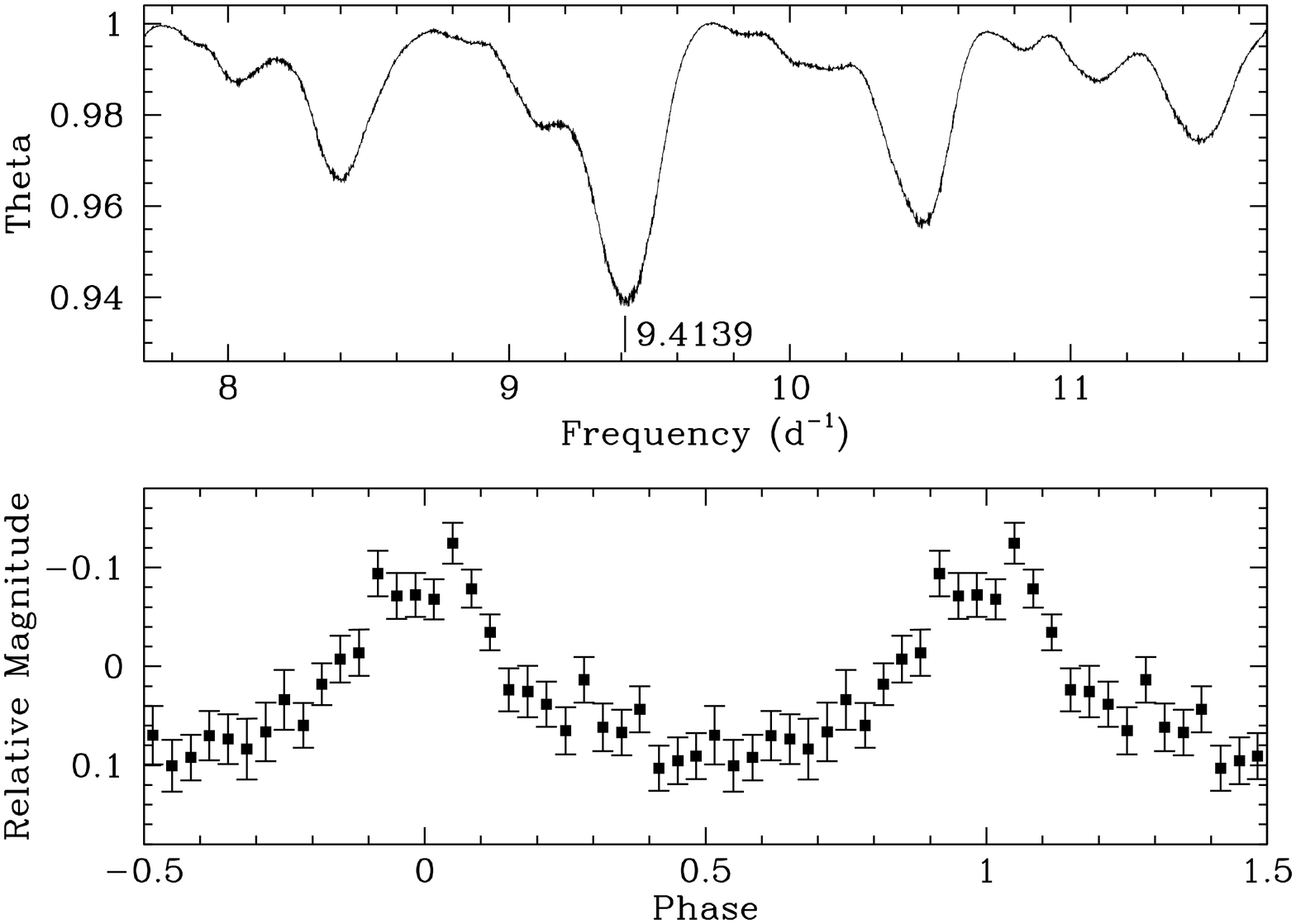}
\caption{{\bf a)} PDM theta diagram for superhumps observed during
 the first superoutburst.  The best estimated superhump period is
 0.104885(93) d.  {\bf b)} Superhump light curve folded by the
 superhump period, after subtracting the mean magnitude from each data
 set.}
\label{fig:super2}
\end{figure}

Fig. \ref{fig:super2} displays the result of the PDM period analysis for
the data obtained during the second superoutburst and the average
superhump light curve.  We used the data marked by $\bigcirc$ in Table
\ref{tab:log} also for this second \Psh\ analysis.  The superhump period
of 0.10623(16) is slightly longer than that during the first
superoutburst.  No apparent signal of a secondary hump around the phase
of 0.5 is seen.

\begin{table}
\caption{Timings of the superhump maxima during the second
 superoutburst.}\label{tab:shmax2}
\begin{center}
\begin{tabular}{crrr}
\hline\hline
HJD$-$2452600 &  E & O$-$C1$^*$ &  O$-$C2$^\dagger$ \\
\hline
17.3991(34) &  0 & $-$0.0525 & $-$0.0094 \\
18.5286(11) & 10 &  0.0002 &  0.0070 \\
19.2918(45) & 17 &  0.0096 &  0.0017 \\
19.2993(40) & 17 &  0.0171 &  0.0092 \\
19.9457(56) & 23 &  0.0174 &  0.0038 \\
20.0461(83) & 24 &  0.0102 & $-$0.0038 \\
20.2696(26) & 26 &  0.0183 &  0.0042 \\
20.3792(39) & 27 &  0.0202 &  0.0063 \\
20.9048(51) & 32 &  0.0074 & $-$0.0027 \\
21.0112(56) & 33 &  0.0061 & $-$0.0027 \\
21.2189(26) & 35 & $-$0.0015 & $-$0.0073 \\
21.3281(38) & 36 &  0.0000 & $-$0.0040 \\
21.3289(62) & 36 &  0.0008 & $-$0.0032 \\
21.9520(84) & 42 & $-$0.0222 & $-$0.0115 \\
22.9117(56) & 51 & $-$0.0316 &  0.0131 \\
\hline
\multicolumn{4}{l}{$^*$ Using Eq. (3).}\\
\multicolumn{4}{l}{$^\dagger$ Using Eq. (4).}\\
\end{tabular}
\end{center}
\end{table}

\begin{figure}
\centering
\includegraphics[width=8.4cm]{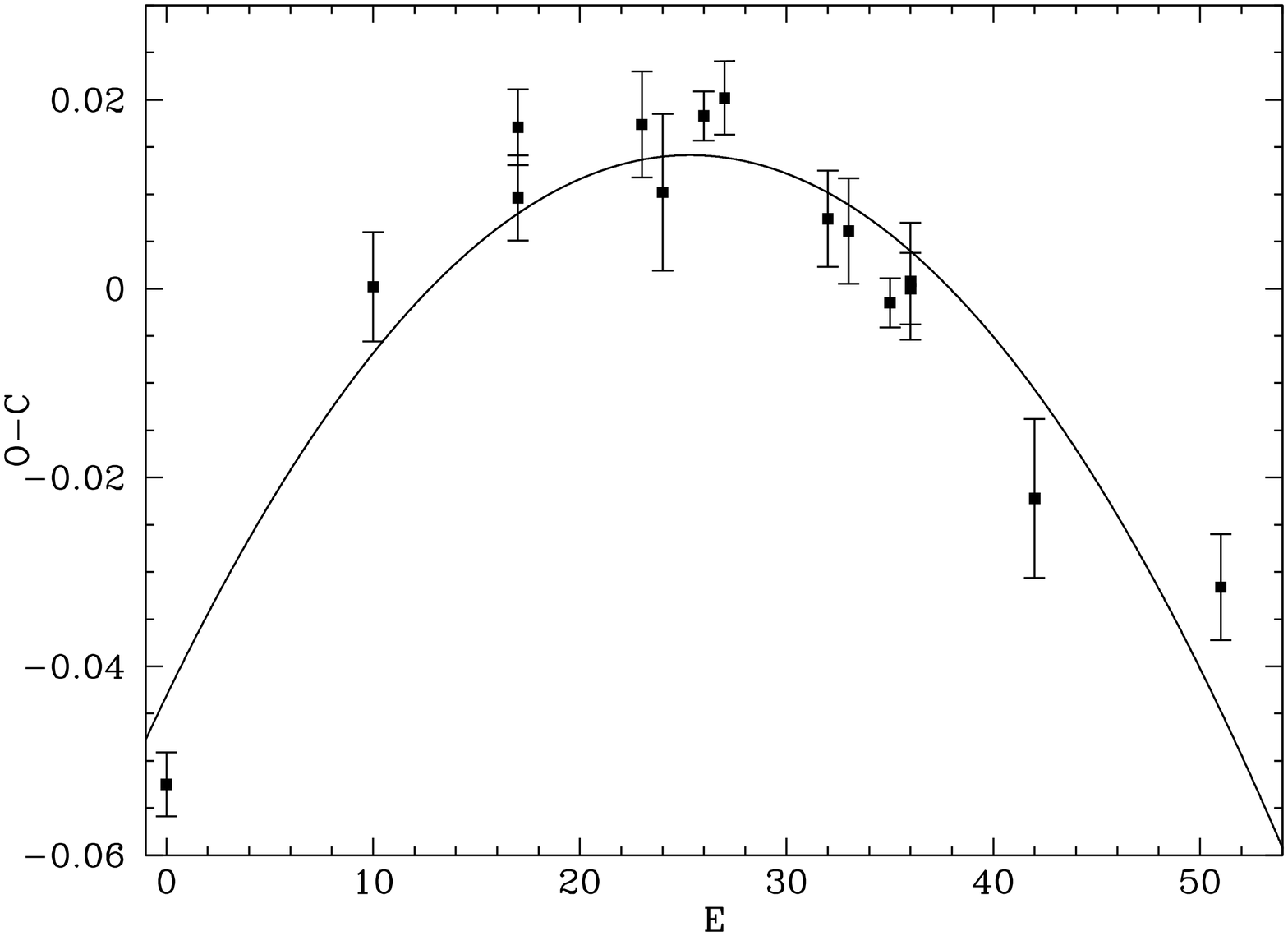}\label{fig:shmax2}
\caption{$O-C$ diagram of the timings of the superhump maxima in Table
 \ref{tab:shmax1}.  The calculated timings are given by Eq. (1).  The
 parabolic curve is based on Eq. (2).}
\end{figure}

The timings of the superhump maxima were obtained for this superoutburst
as before (Table \ref{tab:shmax2}).  A linear regression to these
timings yields the following ephemeris:
\begin{equation}
 HJD_{\rm max} = 20.2513(55) + 0.10768(44)\times(E-26).
\end{equation}
The $O-C1$ calculated using Eq. (3) is displayed in
Fig. \ref{fig:shmax2}.  The diagram clearly shows the decrease in the
superhump period.  Fit to a quadratic equation of the same timings
gives:
\begin{eqnarray}
 HJD_{\rm max} &=& 20.2654(25) + 0.10756(16)\times(E-26) \nonumber \\
               & & - 0.0000893(95)\times(E-26)^2.
\end{eqnarray}
The quadratic term means that the superhump period decreased with a rate
of  \Pdot\ = $\stackrel{.}{P_{\rm SH}}/P_{\rm SH} =
-1.7(2)\times10^{-3}$, which is one order of magnitude larger than the
largest values known (see \cite{kat03v877arakktelpucma}).

\section{Discussion}

\subsection{Two superhump periods}

We obtained two superhump periods: 0.104885(93) d during the first
superoutburst (hereafter \Psh1), and 0.10623(16) d during the second
superoutburst (hereafter \Psh2).  The difference between \Psh1 and \Psh2
must result from difference of the observed phase in the course of
the superoutburst.

The first superoutburst is estimated to have attained to its maximum
brightness at HJD 2452552.0 ($\pm$1.0) from Table \ref{tab:log}.  The
data used for the \Psh1 analysis were therefore taken between the
6($\pm$1)th day and the 14($\pm$1)th day from the supermaximum.  In the
case of the second superoutburst, the maximum of the outburst was
reached somewhen between HJD 2452611.0 and 2452615.0.  Thus the data
used for the \Psh2 analysis were taken between the 4($\pm$2)th day and
the 9($\pm$2)th day from the onset.  Therefore the ``mid'' day of the
observed phase during the second superoutburst (the 7($\pm$2)th day) is
earlier than that during the first superoutburst (the 10($\pm$1) day).
The extremely large change rate observed during the second superoutburst
can easily yield the difference between two superhump periods.

It should be also noted that \Psh\ seemed to decrease with a larger rate
in an earlier phase.  This trend is suggested by the fact that the
change rate of \Psh2 was derived from the superhump-maximum times
between 4($\pm$2)th day and 10($\pm$2)th day, in contrast to that that
of \Psh1 was derived from the timings of the superhump maximum between
9($\pm$1)th day and 10($\pm$1)th day from the supermaximum.

\subsection{Orbital period}

We photometrically detected coherent modulations with a period of
0.10424(3) d in quiescence.  This period is slightly shorter than the
superhump periods \Psh1 and \Psh2, and is naturally attributed to the
orbital period, thus.  Confirmation by spectroscopic observations is,
however, desired, since our quiescence data contain large errors and the
actual error of the period derived is perhaps larger than the noted one
statistically calculated.  The orbital period of 0.10424 d is the second
longest among those of SU UMa stars with the orbital period measured,
next to 0.1172 d of TU Men \citep{men95tumen}, and places Var73 Dra at
the midst of the period gap.

The superhump excess $\epsilon$ $(=(P_{\rm SH}-P_{\rm orb})/P_{\rm
orb})$ is 0.6\% for \Psh1 or 1.9\% for \Psh2, respectively.  It is
generally known that there is a robust relationship that the superhump
excess smoothly increases with \Porb\ (see e.g. \cite{pat98evolution}).
This relationship is well explainable in the disk instability model in
that a large superhump excess suggests a large accretion-disk radius in
a long-\Porb\ system with a large mass ratio ($q = M_{\rm 2}/M_{\rm
1}$).  While Var73 Dra is expected to have $\epsilon \sim$ 5--7\% from
this relation, the derived values of $\epsilon$ corresponds to those of
SU UMa stars with a period about 0.06 d.  This implies that Var73 Dra
has a small mass ratio, although theoretical calculations on the CV
evolution propose a high mass ratio for a CV in the period gap
(e.g. \cite{how01periodgap}).  Var73 Dra may be the first object which
breaks the $\epsilon$--\Porb\ relation[\citet{pat98evolution} discusses
this relationship after correction of \Psh, taking period changes into
account, to the value 4 days after superhump emergence.  The same
correction does not have significant effect on our results.].  This
problem urges spectroscopic determination of $q$ as well as \Porb.  Note
that the modulations in quiescence may be attributed to permanent
superhumps, as discussed later.

\subsection{Derivative of the superhump period}

Until mid 1990s, the superhump period was considered to monotonically
decrease or at least be constant after full development (see
e.g. \cite{war85suuma,pat93vyaqr}).  This phenomenon was basically
explained in the disk instability scheme by that the precession
frequency of the eccentric accretion disk decreases due to shrinkage of
the disk radius \citet{osa85SHexcess}, or propagation of the eccentric
wave to the inner disk \citet{lub92SH}.  Elongation of \Psh\ was,
however, first observed during the 1995 superoutburst of AL Com
\citep{nog97alcom}.  Following this discovery, similar behavior has been
found in several SU UMa stars: V485 Cen \citep{ole97v485cen}, EG Cnc
\citep{kat97egcnc}, SW UMa \citep{sem97swuma, nog98swuma}, V1028 Cyg
\citep{bab00v1028cyg}, WX Cet \citep{kat01wxcet}, and HV Vir
\citep{kat01hvvir}.  These stars are, however, concentrated around the
period minimum in the \Porb distribution, and SU UMa stars with
relatively long \Porb have been confirmed to show \Psh\ decrease with a
similar rate of \Pdot\ $\sim -5 \times 10^{-5}$ (see
\cite{kat03v877arakktelpucma}).  Very recently,
\citet{kat03v877arakktelpucma} reported large negative derivatives of
\Psh\ in V877 Ara (\Pdot\ = $-$1.5($\pm$0.2) $\times$ 10$^{-4}$, \Psh\ =
0.08411(2) d) and KK Tel (\Pdot\ = $-$3.7($\pm$0.4) $\times$ 10$^{-4}$,
\Psh\ = 0.08808 d), and pointed out a diversity of \Pdot\ in long-period
SU UMa-type dwarf novae.

We revealed that \Pdot\ in Var73 Dra during the second superoutburst was
still about one order of magnitude larger than these two records.
\citet{kat01wxcet} and \citet{kat03v877arakktelpucma} proposed a
possibility that \Pdot is related to the mass transfer rate: SU UMa
stars with larger \Pdot\ tend to have larger mass transfer rates, and
those with \Pdot\ close to and smaller than zero have small \Mdot.  The
quite short supercycle length of about 60 d suggests a high \Mdot\ in
the present object (discussed later), which may support this possiblity.
It should be, however, worth noting that \citet{kat03bfara} found \Pdot\
$\sim0$ in BF Ara, an SU UMa star supposed to have a rather large \Mdot.

\subsection{Outburst behavior}

The three superoutbursts we caught suggests that Var73 Dra steadily
repeats superoutbursts with a supercycle of $\sim$60 d.  This value is
shorter than the shorterst one known so far in usual SU UMa stars (89.4
d in BF Ara: \cite{kat03bfara}), and close to 19--50 d of ER UMa stars.

The disk instability model predicts that the supercycle is shorter in an
SU UMa-type star with a highter \Mdot.  Reproduction of the light curves
of ER UM stars was successfully done by \citet{osa95eruma} by assuming a
mass transfer rate about ten times higher than that in ordinary SU UMa
stars (see also \cite{osa95rzlmi}), although it has not still been
clear why ER UMa stars have such high mass transfer rate
\citep{nog98CVevolution}.  Var73 Dra is expected to also have a very
high mass transfer rate because of its extraordinary short supercycle
\citep{ich94cycle}.  This condition may be achieved if this star is in
the short, high-\Mdot\ phase just after getting semi-detached and
starting mass transfer.  This interpretation provide an explanation to
the problem on the evolutionary status of this star that mass transfer
is supposed to be stopped (or seriously reduced) for evolution in the
period gap in the currently standard evolution theory.

This simple view, however, faces a difficulty of lack of the normal
outburst in Var73 Dra.  We caught two superoutbursts and two normal
outbursts in the course of monitoring.  The recurrence cycle of the
normal outburst and the supercycle are estimated to be over 15 days and
$\sim$60 days, respectively.  In contrast, the normal-outburst
recurrence cycle is expected to be $\sim$8 days for an SU UMa star with
a supercycle of 60 days based on the model reviewed by
\citet{osa96review}.

The normal-outburst cycle was, however, 7--8 d by \citet{ant02var73dra}
from their observations in 2001 August--October.  The supercycle at that
time was longer than at least 70 d, judging from Fig. 3 in
\citet{ant02var73dra}.  These facts clearly indicate a chage of the
outburst activity between 2001 and 2002.  Similar changes have been
reported in recent years, such as in DI UMa \citep{fri99diuma}, SU UMa
(\cite{ros00suuma, kat02suuma}), V1113 Cyg \citep{kat01v1113cyg}, V503
Cyg \citep{kat02v503cyg}, and DM Lyr \citep{nog03dmlyr}.  Among thse
stars, only DM Lyr showed an anti-correlation: the recurrence cycle of
the normal outburst decreased, and the supercycle increased, while Var73
Dra showed a reverse anti-correlation: the recurrence cycle of the
normal outburst increased, and the supercycle decreased.  Such
behavior can not be explained by variation of the mass transfer rate
due to e.g. the solar-type cycle of the secondary star
(e.g. \cite{ak01CVcycle}).  \citet{nog03dmlyr} proposed for DM Lyr that
a machanism to reduce the number of the normal outbursts may work when
the superoutbursts more frequently occur and another mechanism to
shorten the recurrence time of the normal outburst may work when the
superoutburst less frequently takes place.  The same idea may be
applicable to Var73 Dra.  Closer monitoring to avoid to miss rather
faint normal outbursts ($>$15 mag) is needed to check variabilities
of the recurrence cycles of the normal outburst and superoutburst.

\subsection{Related to ER UMa stars?}

Two problems regarding ER UMa stars to be solved are the extraordinary
large mass transfer rates for their short orbital periods and the
evolution path, as mentioned above.  One of the keys to the problems is
discovery of ER UMa counterparts with longer \Porb.

Whether Var73 Dra is an object related to ER UMa stars is in interesting
subject.  While the supercycle of $\sim$60 d is certainly very close to
those of ER UMa stars, our observations give a negative support to this
question in terms of the duty cycle of the superoutburst and the
recurrence cycle of the normal outburst.  The duration of the supercycle
of Var73 Dra is at most 15 d (Table \ref{tab:log}), a normal one for an
SU UMa system, and the duty cycle of the superoutburst in one supercycle
is $\sim$25\%, while the duty cycle is 30--50\% in ER UMa stars.  The
normal outburst is 1 or at most a few in one supercycle, quite
infrequent for an ER UMa analog.

New interpretations on how ER UMa stars most frequently give rise to
superoutbursts have been recently published, which are based on the disk
instability scheme, but assuming decoupling of the thermal and tidal
instability \citep{hel01eruma}, or the effects of irradiation
\citep{bua02suumamodel}.  Both models predict superhumps observed in
quiescence.  The modulations observed here in quiescence may be
superhumps, which could give a solution to the problem that the
superhump excess in Var73 Dra is too small for this long \Porb.  A small
mass ratio is, however, a basic assumption in both models.  Measurement
of the orbital period and the mass ratio in this system has a
significant effect also on this matter.

\citet{kat03erumaSH} discovered a peculiar behavior of superhumps in ER
UMa which is a phase shift of 0.5 before entering the plateau phase of
the superoutburst, and interpreted that the (normal) superhumps are seen
at the very early phase of the superoutburst and the modulations
observed during the plateau phase correspond to `late' superhumps in SU
UMa stars.  To trace the superhump evolution throughout a superoutburst
is important to clarify the superhumps in Var73 Dra exhibit the normal
SU UMa-type behavior or the ER UMa-type one.

\begin{acknowledgements}

 This research has made use of the USNOFS Image and Catalogue Archive
 operated by the United States Naval Observatory, Flagstaff Station
 (http://www.nofs.navy.mil/data/fchpix/), and the SIMBAD database,
 operated at CDS, Strasbourg, France
 (http://simbad.u-strasbg.fr/Simbad).  GM acknowledges the support of
 Software Bisque and Santa Barbara Instrument Group.  This work is
 partly supported by a grant-in aid (13640239) from the Japanese
 Ministry of Education, Culture, Sports, Science and Technology (TK),
 and by a Research Fellowship of the Japan Society for the Promotion of
 Science for Young Scientists (MU).

\end{acknowledgements}

\end{document}